\documentclass[12pt]{iopart}
\usepackage{amsfonts}
\begin{document}
%\parindent 0 pt
%\renewcommand{\baselinestretch}{1.5}

%%%%%%%%%%%%%%%%%%%%%%%%%%%%%%%%%

\def \ep{\epsilon} 
\def \intR {\int_{-\infty}^{+\infty}}  
\def \grad {\nabla}  
\def \ov{\over} 
\def \q  {\quad}  
\def \qq {\qquad}  
\def \bar{\overline} 
\def \beq{ \begin{equation} } 
\def \eeq{\end{equation}} 
\def \r#1{$^{#1}$}
\newtheorem{corollary}{Corollary}
\newtheorem{proposition}{Proposition}
\newtheorem{lemma}{Lemma}                 
\newtheorem{theorem}{Theorem}
\renewcommand{\labelenumi}{(\roman{enumi})}

%%%%%%%%%%%%%%%%%%%%%%%%%%%%%%%%%%%%%%%%%%%%%%%%%%%%%%%%%%%%%%%%%%%%%%%%%%%

\title[Geometrical Obstruction to the Integrability of Geodesic Flows]{Geometrical
Obstruction to the Integrability of Geodesic Flows: an Application 
to the Anisotropic Kepler Problem}

\author{Manuele Santoprete\dag }

\address{\dag Department of  Mathematics and Statistics  University of 
 Victoria, P.O.  Box  3045 Victoria B.C.,  Canada, V8W 3P4}

\ead{msantopr@math.uvic.ca}

 %\footnotetext[1]{Electronic mail: msantopr@math.uvic.ca}
\begin{abstract}
\noindent Resorting  to classical techniques of  Riemannian geometry we
develop   a geometrical   method   suitable   to   investigate   the
nonintegrability of  geodesic flows and of natural Hamiltonian systems.  Then we apply  such method
 to  the Anisotropic Kepler  Problem (AKP) and we prove that it is not analytically integrable.
% This application   is  particularly interesting since  the  nonintegrability 
%(and the occurrence of chaos), for weak anisotropies,  is an outstanding problem.  
\end{abstract}

%\submitto{\NL}  

%\pacs{}
%37J30= Obstruction to integrability
%37J30 Symmetries, invariants, invariant manifolds, momentum maps, reduction.

\maketitle
%\small\normalsize

%%%%%%%%%%%%%%%%%%%%%%%%%%%%%%%%%%%%%%%%%%%%%%%%%%%%%%%%%%%%%%%%%%%%%%%
\section{Introduction}
%%%%%%%%%%%%%%%%%%%%%%%%%%%%%%%%%%%%%%%%%%%%%%%%%%%%%%%%%%%%%%%%%%%%%%%

 Let $M$ be a  manifold and let $TM$ and $T^\ast M$  be the tangent and
 cotangent bundles  of $M$, respectively.  Moreover let  $g_{ij}(x)$ be a
 Riemannian metric on $M$. The  corresponding flow on $TM$ is given by
 the Lagrangian ${\cal  L}=1/2g_{ij}(x)x^1x^j$. The resulting Hamiltonian
 system  on  $T^\ast  M$  has  Hamiltonian  $H=1/2g^{ij}(x)p_ip_j$.

 Such systems are very important examples of Hamiltonian systems, since many
 problems  in mechanics  reduce  to  the study  of  geodesic flows  on
 Riemann and  Finsler metrics \cite{Kozlov2} . For  example, with a
 suitable  transformation, the Kepler  problem can  be described  as the
 geodesic  flow  on a  sphere  \cite{Moser,Osipov}.  

 In  this paper  we  are
 interested  in  developing  a  geometrical  method  to  investigate whether 
  systems that can  be reduced to geodesic flows on
 a manifold are integrable. 

We recall that an Hamiltonian system with $n$ degrees of
 freedom  is   {\it integrable}  if  there  exist   $n$  first  integrals
 $I_1,\dots,I_n$ with  the property  that $\{I_i,I_j\}=0$ for  all $i$
 and $j$, and such that they  are functionally independent.
In particular the definition above can be used when the Hamiltonian describes
 the geodesic flow on a manifold.

If $M$  is an analytic manifold  and the metric  $g_{i,j}$ is analytic
then  the flow  is said  to be  {\it analytically  integrable}  if the
indicated integrals  $I_1,\cdots,I_n$ exist and  can be chosen  to be
analytic.

In  \cite{Kozlov2,Kozlov} it  was proved that  the geodesic
flow of  any analytic  metric on an  analytic compact  two dimensional
manifold  of  genus  $g>1$  is  not  analytically  integrable  and  in
\cite{Taimanov}  the  discussion was  generalized  to $n$  dimensional
manifolds.  However it is clear that the obstructions to integrability
are  not  only  of  topological  origin.  This  is  because  different
Riemannian metrics can  be associated with the same  manifold, even if
the topology  (of the manifold)  imposes restrictions to  the possible
metrics.    Indeed   Donnay  \cite{Donnay}   and   Burns  and   Gerber
\cite{Burns}  have constructed  real  analytic metrics  on the  sphere
whose geodesic flows are Bernoulli. Knieper and Weiss \cite{Weiss} constructed 
real analytic metrics with positive curvature and positive topological entropy.
 Thus, even if the sphere has genus
$g=0$ they  found that there are  metrics on the sphere  such that the
geodesic  flow is not  analytically integrable.   Hence to  refine the
criterion  of  Kozlov  (and  Taimanov)  it  is  necessary  to  find  a
geometrical  approach  that takes  into  account  the peculiar  metric
defined  on the  manifold.     Moreover the topological considerations
of Kozlov and Taimanov can be applied only to compact manifolds and it would
be useful to find a technique applicable in the non compact case.
 The main  purpose  of this  paper is  to consider  such  geometrical  approach
that both takes into account the peculiar metric and can be applied to
non-compact manifolds.

The approach  we consider lies on the  simple idea that if
you have an integrable system and you add a perturbation that destroys
the initial symmetry then, unless  a new symmetry appears , you obtain
a non integrable system. Examples of this situations are discussed in
\cite{Diacu} and \cite{Santoprete}, where suitable generalizations of
the  Poicar\'e-Melnikov   method  (see  for   example  \cite{Cicogna1,Cicogna2}
and references therein)  are applied to the perturbed systems.
In \cite{Diacu} the Anisotropic Manev problem is discussed for weak anisotropies,
and resorting to the Melnikov method the occurrence of chaos on the zero 
energy manifold is proven. In \cite{Diacu} the broken symmetry is the 
rotational invariance. Similarly in \cite{Santoprete}, where we discussed
 black holes immersed in uniform electric and magnetic fields, the broken symmetry
was the rotational invariance.    

To formalize this intuitive idea, in the next section some results of 
Riemannian geometry are recalled, in particular the isometries, the 
infinitesimal isometries and some of their properties  are  discussed.
Then we observe that the equation of Killing can be used to find the symmetries
that preserve the equation of motion and   hence if those equation have only
 the zero solution no first integral can be found other than the Hamiltonian.

In Section 3 we introduce our main example: the Anisotropic Kepler Problem (AKP) and 
we show that with a suitable canonical transformation it can be reduced to
the geodesic flow on a Riemannian surface with non constant curvature.
The Anisotropic Kepler Problem describes the motion of one body  orbiting 
around another considered fixed at the origin in an anisotropic space 
(i.e. such that the force acts differently in each direction). 
The AKP was carefully studied in a series of paper by various authors
(see \cite{Casasayas,Gutzwiller}  for an overview of the problem  and of the main results),
but not  much is known for weak anisotropies. Indeed even if there was evidence
of a chaotic behaviour no proof of this fact can be found in the literature (except for the 
occurrence of chaos on the zero energy level that was proved in \cite{Diacu}).
Moreover the AKP is an example of geodesic flow on a non-compact
surface and hence it is particularly suitable to illustrate some of
the peculiarities of the technique under discussion in this work.
Hence showing the nonintegrability of the AKP for weak anisotropies seems an interesting
application.

The last section is devoted to show that the Equations of Killing do not have any non trivial solutions for 
the AKP and hence that, by the results of Section 3, the Anisotropic Kepler Problem has no  analytic first  integral
independent of the Hamiltonian.

%%%%%%%%%%%%%%%%%%%%%%%%%%%%%%%%%%%%%%%%%%%%%%%%%%%%%%%%%%%%%%%%%%%%%%%
\section{Isometries and Infinitesimal Isometries: The Killing's Field}
%%%%%%%%%%%%%%%%%%%%%%%%%%%%%%%%%%%%%%%%%%%%%%%%%%%%%%%%%%%%%%%%%%%%%%%
We will now  recall  some  results and  known  facts from  Riemannian
geometry that we need in order  to prove the main result of the paper;
for    a    more     detailed    discussion    see,    for    example,
\cite{Dubrovin,Kobayashi,Eisenhart} .
 
Let  $M$  be  a $n$-dimensional  manifold  with  a  Riemannian  metric
$g_{ij}$. An isometry  of $M$ is a transformation  of $M$ which leaves
the   metric  invariant.   More  precisely,   given   the  coordinates
$x_1,\dots,  x_n$ on  $M$, a  transformation $x_i=x_i(\bar  x_1, \dots,
\bar  x_n)$ is  an isometry  (or motion  of the  metric) if
\beq \bar
g_{ij}(\bar x_1, \dots, \bar x_n)= g_{ij}(x_1(\bar x), \dots, x_n(\bar
x)).  \eeq
It is clear that such motions  form a group, that is often called group
of motion. 
Moreover a vector field $\xi^i$ is called an infinitesimal
isometry (or a Killing vector field) if the local 1-parameter group of
local transformations generated by  $\xi^i$ in a neighbourhood of each
point of $M$  consists of local isometries.  The  set of infinitesimal
isometries  of $M$, denoted  by $i(M)$,  forms a  Lie algebra.   Now a
transformation is an infinitesimal motions  of $M$ into itself, i.e. a
transformation  that  preserves  the   metric  if  the  Lie  derivative
vanishes, that is, if the equations of Killing

\beq
L_{\xi}g_{ij}=\xi^k {\partial g_{ij}\over \partial x^k} +g_{ik}{\partial \xi^k\over\partial x^j}+g_{jk}
{\partial \xi^k \over \partial x^i}=0
\label{killing}
\eeq
are satisfied \cite{Eisenhart}. In order that there may be a non-null 
motion we must have the additional condition
\beq
g_{ij}\xi^i\xi^j \neq 0.
\eeq 
The first fact  to verify
is that the infinitesimal  isometries are all the transformations that
preserve   the   variational  problem   defined   by  the   Lagrangian
${\cal L}=1/2g_{ij}x^ix^j$, so that we can  reduce the study of the symmetries
preserving the equation of motion to   the study  of the  isometries. This
problem is solved by the following
\begin{lemma}
An  infinitesimal transformation  is an  isometry for $M$ with the
metric ${g_{ij}}$  if and  only if  it
preserves  the variational  problem defined  by the  Lagrangian ${\cal
L}=1/2 g_{ij}\dot x^i \dot x^j$ on $TM$.
\label{3}
\end{lemma}
PROOF Let us remark that a transformation is an infinitesimal isometry
 if and only if the Lie derivative  of the metric is zero, i.e. if the
 Killing's equations (\ref{killing}) are satisfied.  On the other hand
 a  transformation preserves  the variational  problem defined  by the
 Lagrangian (\ref{lagrangian}) when the Lie derivative with respect to
 the field $\xi^i$ is zero,
\beq
L_{\xi}({\cal L})=\xi^i {\partial {\cal L}\over \partial x^i}+
 {\partial \xi^i \over \partial x^j} \dot x^j {\partial {\cal L}\over \partial \dot x^i}=0
\eeq
 that after few manipulations takes the form
\beq
L_{\xi}({\cal L})={1 \over 2} \left\{ \xi^s {\partial g_{ij} \over \partial x^s}+
{\partial \xi^k \over \partial x^i} g_{kj}+
{\partial \xi^k \over \partial x^j} g_{ik} \right\} \dot x^j \dot x^i=0 . 
\label{Lie}
\eeq   It   is   trivial   to   see   that   (\ref{killing})   implies
(\ref{Lie}). The converse is also true since $x^i$ is arbitrary and the tensor in the braces
 is symmetric. This concludes the proof.

With the preparations above we are well on our way to establishing the following

\begin{theorem}
Let $M$ be a analytic  $n$-manifold with Riemannian metric $g_{ij}$ and  ${\cal L}=1/2g_{ij}x^ix^j$
 be the Lagrangian defined on $TM$.  If the Killing's equations for
 $M$ have less then $n-1$
 independent solutions the geodesic flow is not analytically integrable.
\end{theorem}  
To complete the proof of the theorem we can remark that to each integral of 
motion there corresponds a one parameter group of transformation that preserve the 
equation of motion \cite{Arnold}. Hence, since the equation of Killing admit less than $n-1$ independent 
 solutions,  by Lemma \ref{3}, there are less than $n-1$ transformations that preserve the variational problem.
The remark above concludes the proof.

\noindent From the previous theorem we easily obtain the following

\begin{corollary}
If the only solution of the Killing's equations is $\xi^i\equiv 0$, the corresponding
 geodesic flow is not analytically integrable.
\end{corollary}
We  now want  to consider  some more  properties of  the infinitesimal
isometries. A classical  result \cite{Kobayashi,Eisenhart}  is the
following
\begin{proposition}
The  Lie algebra  of infinitesimal  isometries $i(M)$  of  a connected
Riemannian manifold $M$  is of dimension at most  ${1\over 2} n(n+1)$,
where $n=dim  M$. If  $dim~ i(M)=  {1\over 2} n(n+1)$,  then $M$  is a
space of constant curvature.
\end{proposition}
This Proposition  gives  the maximum number  of infinitesimal isometries  and hence
 also the maximum  number of first integrals for a  geodesic flow on a
 manifold  of dimension  $n$. In particular in the two dimensional case we have
\begin{corollary}
 If  $M$ is  a two  dimensional connected  Riemannian manifold  of non
constant curvature  then the  Lie algebra of  infinitesimal isometries
$i(M)$ is of dimension at most $1$.
\label{4}
\end{corollary}
One example that illustrates the results above is the 
 Kepler problem on the plane. Such problem can  be described as the geodesic flow on
 a  sphere $S^2$,  hence the  isometries form  a Lie  algebra 
 isomorphic to $SO(3)$, and there are three first integrals other than
 the  Hamiltonian  (one component  of  the  angular  momentum and  two
 components of the Runge-Lenz vector). On the other hand, if we consider,
for instance, an homogeneous potential, other than the harmonic oscillator 
and the Kepler one, we can describe the motion only as a geodesic flow on a 
surface that has a $SO(2)$ symmetry.

%%%%%%%%%%%%%%%%%%%%%%%%%%%%%%%%%%%%%%%%%%%%%%%%%%%%%%%%%%%%%%%%%%%%%%%
\section{The Anisotropic Problems as Geodesic Flow on a Surface}
%%%%%%%%%%%%%%%%%%%%%%%%%%%%%%%%%%%%%%%%%%%%%%%%%%%%%%%%%%%%%%%%%%%%%%%

The Hamiltonian function for the Anisotropic Kepler Problem on the plane can be written as

\beq 
H({\bf q},{\bf p})={{\bf p}^2 \over 2}-{a \over (q_1^2+\mu q_2^2)^{1/2}} 
\label{anisotropic}
\eeq
where $\mu>1$ is a constant and ${\bf q}=(q_1,q_2)$ is the position of
one body with respect to the other, considered fixed at the origin of the
 coordinate system, and ${\bf p}=(p_1,p_2)$ is the momentum of the moving particle. 
The constant measures the strength of the anisotropy and for $\mu=1$ we recover the 
classical Kepler Problem. 

With  some  canonical transformations we  want to write the  problem as
the geodesic  flow on a  surface.  To accomplish this task we  need the
following Lemma 

\begin{lemma}
Let $({\bf q}(t),{\bf p} (t))$ be a solution of the Hamiltonian system with Hamiltonian $ H({\bf q},\bf{p})$
which lies on the level surface $H=0$.
Let us change the time variable $t$  to $\tau$ along this trajectory according
to     the    formula     $d\tau/dt=1/G({\bf q}(t),{\bf p}(t))\neq 0$.    Then
$({\bf q}(\tau),{\bf p}(\tau))=({\bf q}(t(\tau)),{\bf p}(t(\tau)))$  is   a  solution  of  the
Hamiltonian system (with respect  to the  same symplectic  structure) 
with    Hamiltonian   $\tilde    H=HG$.   If    $G=2(H+\alpha)$   with
$\alpha=\mbox{const}$, one can also put $\tilde H=(H+\alpha)^2$.
\label{1}
\end{lemma}
The statement above can be found in \cite{Arnold} and the proof is
based on the idea of extended phase space (see \cite{Dubrovin}).
If we perform the  change of time $d\tau/dt=(q_1^2+\mu q_2^2)^{1/2}$
on the submanifold $H=h$, according to Lemma \ref{1}, we obtain  the new Hamiltonian

\beq
(q_1^2+\mu q_2^2)^{1/2}(H-h)=(q_1^2+\mu q_2^2)^{1/2}\left({{\bf p}^2-2h \over 2}  \right)-a.
\eeq
Then again change the time $\tau$ to $s$ by $ds/d\tau=2((q_1^2+\mu q_2^2)^{1/2}(H-h)+a)$ 
on the same level surface $H=h$. By the second part of Lemma \ref{1} we obtain the Hamiltonian system with Hamiltonian
\beq
\tilde H=(q_1^2+\mu q_2^2){({\bf p}^2-2h)^2 \over 4}.
\label{hamiltonian}
\eeq
Finally, applying Legendre's transformation, regarding ${\bf p}$ as coordinates and ${\bf q}$ 
as the canonically conjugate momenta, we can write

\beq \left \{\begin{array}{l}
p_1'={\partial \tilde H \over \partial q_1}={q_1 \over 2}({\bf p}^2-2h)^2 \medskip\\
p_2'={\partial \tilde H \over \partial q_2}={q_2 \mu \over 2}({\bf p}^2-2h)^2
\end{array} \right .
\eeq
where the prime indicates the derivative with respect to $s$.
Consequently

\beq \left \{ \begin{array}{l}
q_1={2p_1'\over ({\bf p}^2-2h)^2} \medskip \\
q_2={2p_2'\over \mu({\bf p}^2-2h)^2} 
\end{array} \right . .
\eeq
Moreover since we have that
\beq 
{\cal L}=p_1'q_1({\bf p},p_1')+p_2'q_2({\bf p},p_2')-\tilde H(q_1({\bf p}, p_1'),q_2({\bf p}, p_2'),p_1,p_2 )= 
\eeq
we obtain a natural Lagrangian system with Lagrangian
\beq
 {\cal L} = {1 \over ({\bf p}^2-2h)^2}\left(p_1'^2+{p_2'^2\over \mu}\right)
\eeq
and if we let $p_1=x, p_2=y$ and $x^i=(x,y)$ we obtain
\beq
{\cal L}={1 \over 2} g_{ij}\dot x^i\dot x^j
\label{lagrangian}
\eeq
where 
\beq
g_{ij}={2 \over (x^2+y^2-2h)^2}\left(
\begin{array}{cc}
1 & 0 \\
0 & 1/\mu
\end{array}
\right).
\label{metric}
\eeq
The Lagrangian above defines a Riemann metric of Gaussian curvature given by
\beq
K= -(1-\mu)(x^2-y^2)-2h(1+\mu)
\label {curvature}
\eeq
 that for $\mu=1$ describes a metric of constant 
 Gaussian curvature (positive for $h<0$ and negative for $h>0$)  \cite{Moser, Osipov,Arnold}.
It is interesting to remark that, by Corollary 2, for $\mu /neq 1$ we have at most one infinitesimal 
isometry (and hence also at most one first integral), since the curvature is not constant.

%%%%%%%%%%%%%%%%%%%%%%%%%%%%%%%%%%%%%%%%%%%%%%%%%%%%%%%%%%%%%%%%%%%%%%%
\section{Nonexistence of Solutions of the Killing's Equations And Nonintegrability in the Anisotropic Problem}
%%%%%%%%%%%%%%%%%%%%%%%%%%%%%%%%%%%%%%%%%%%%%%%%%%%%%%%%%%%%%%%%%%%%%%%

In the previous section we reduced the Anisotropic Kepler Problem to a geodesic flow
 on a two dimensional surface, and hence we can apply the results obtained in Section 2. This section
is devoted to prove the nonintegrability of such system.
Thus we are well on our way to proving the following
\begin{theorem}
The Anisotropic Kepler Problem is not analytically integrable, i.e. it does not have any analytic 
first integral independent of the Hamiltonian.
\end{theorem}

The only fact that remains to be proved is that the Killing's equations doesn't have any non-null solution.
The equations of Killing for the metric (\ref{metric}) can be written as 
\beq
\left \{\begin{array}{c}
{\partial \xi^1 \over \partial y}+ {1 \over \mu}{\partial \xi^2 \over \partial x}=0 \medskip\\
-4{(x\xi^1+y\xi^2) \over (x^2+y^2-2h)}+2{\partial \xi^1 \over \partial x}=0\medskip\\
-4{(x\xi^1+y\xi^2) \over (x^2+y^2-2h)}+2{\partial \xi^2 \over \partial y}=0
\end{array} \right .
\label{system}
\eeq
Subtracting the third equation from the second we get
\beq
{\partial \xi^1 \over \partial x}={\partial \xi^2\over\partial y},
\label{partials}
\eeq
combining  (\ref{partials}) with the first equation in system (\ref{system}) we obtain

\beq
{\partial^2 \xi^1 \over \partial x^2}=-\mu{\partial ^2 \xi^1 \over \partial y^2} \q , \q
{\partial^2 \xi^2 \over \partial y^2}=-{1 \over \mu}{\partial ^2 \xi^2 \over \partial x^2}.
\label{secondpartials}
\eeq

Now  multiplying  the  second  equation in  system  (\ref{system})  by
$(x^2+y^2 -2h)$  (assume $h<0$),  differentiating with respect  to $x$
and  using  equation  (\ref{partials})   and  the  first  relation  in
(\ref{system}) we get the following second order differential equation

\beq  2\mu(x^2+y^2 -2h){{\partial^2  \xi^1  \over \partial  y^2}}-4\mu
y{\partial \xi^1 \over \partial y}+4\xi^1=0. \label{eq1} \eeq
Similarly  from the third  equation in  (\ref{system}) we  get another
second order differential equation

\beq  {2 \over \mu}(x^2+y^2 -2h){{\partial^2  \xi^2  \over \partial  x^2}}-{4\over \mu}
x{\partial \xi^2 \over \partial x}+4\xi^2=0. \label{eq2} \eeq
It is easy to  see that if $h<0$, $y=0$ is an  ordinary point since the
functions $-2y/(x^2+y^2-2h)$  and $2/(\mu(x^2+y^2-2h))$ are analytical
in $y$  at $y=0$.  Let each  of these functions be  represented by its
Taylor  series at  $y=0$ on  the real  line ,  then every  solution of
(\ref{eq1}) on $\mathbb{R}$ is analytic at  $y=0$ and can be represented on $\mathbb{R}$
by  its   Taylor  series  at  $y=0$.   Similarly   every  solution  of
(\ref{eq2}) can  be represented on $\mathbb{R}$  by its Taylor  series at $x=0$
since $x=0$ is an ordinary point.
Therefore we  can look for  solutions of equations (\ref{eq1})  of the
form   $\xi^1=\sum_{n=0}^{\infty}a_n(x)y^n    $   and   solutions   of
(\ref{eq2}) of the form $\xi^2=\sum_{n=0}^{\infty}b_n(y)x^n $. Routine
calculations show that, from (\ref{eq1}) we must have

\beq
a_2=-{a_0 \over\mu(x^2-2h)} \q ,\q a_3={(\mu-1)a_1 \over3\mu(x^2-2h)}
\label{a_2}
\eeq
and the following recurrence relation
\beq
a_{n+2}=-{ 2 +\mu n(n-3) \over \mu(x^2-2h)(n+2)(n+1) }a_n
\eeq
holds for $n>1$.
From equation (\ref{eq2}) we get
\beq
b_2=-{\mu b_0 \over(y^2-2h)  } \q ,\q b_3=-{(\mu-1) b_1 \over3(y^2-2h)} 
\eeq
and the recurrence relation 
\beq
b_{n+2}=-{ 2 \mu + n(n-3) \over (y^2-2h)(n+2)(n+1) }b_n
\eeq
for $n>1$.
With the preparations above  we can find  the general solution of (\ref{eq1}) 
\beq \begin{array}{l}
\xi^1=a_0(x)\left(1-{1 \over \mu(x^2-2h)}y^2+{(1-\mu)\over 6\mu^2(x^2-2h)^2}y^4+ \dots \right)+\medskip\\
 \qq + a_1(x)\left( y-{\mu-1 \over 3\mu(x^2-2h)}y^3+{\mu-1 \over 30 \mu^2(x^2-2h)^2}y^5+ \dots\right)
\end{array}
\eeq
and the general solution of (\ref{eq2}) 
\beq
\begin{array}{l}
\xi^2=b_0(y)\left(1-{\mu \over (x^2-2h)}x^2+{(\mu-1)\mu\over 6(x^2-2h)^2}x^4+ \dots \right)+\medskip\\
 \qq + b_1(y)\left( x-{\mu-1 \over 3(x^2-2h)}x^3+{(\mu-1)\mu \over 30 (x^2-2h)^2}x^5+ \dots\right)
\end{array}.
\eeq We will not need the  explicit form of the solutions, but we will
retain  the recurrence  relations.   Now  we need  to  check that  the
solutions  we  found   above  satisfy  the  ``consistency''  relations
(\ref{secondpartials}).
Consider a solution of (\ref{eq1}) written as a power series $\sum_{n=0}^{\infty}a_n(x)y^2$,
 using the first of equations (\ref{secondpartials}) we find the relations
\beq
{d^2a_n \over dx^2} =-2\mu a_{n+2}
\label{miracle}
\eeq
since, by analyticity, we can exchange the sums with the derivatives. It is easy
 to check that choosing  $n=0$ the previous relation reduces to
\beq
{d^2a_0 \over dx^2} =-2\mu a_{2}={2a_0 \over (x^2-2h)}
\label{recurrence3}
\eeq
that is solved by $a_0=(x^2-2h)$.
For $n=2$ instead we have
\beq
{d^2a_2 \over dx^2} =-2\mu a_{4}={1-\mu \over 3(x^2-2h)}a_2.
\label{finally}
\eeq 
From the first  of equations (\ref{a_2}) with $a_0=(x^2-2h)$ we
get  that $a_2=-1/\mu$  doesn't  satisfy (\ref{finally})  if $\mu
\neq 1$. Thus the consistency conditions impose that $a_0=0$ and consequently
$a_{2n}=0$ for every  integer $n$. 

Now, to show that  the odd terms of
the series also vanish, we consider equation (\ref{miracle}) with $n=1$
and  we get  \beq  {d^2a_1 \over  dx^2}  =2\mu a_{3}={2(\mu-1)  \over3
(x^2-2h)}a_1
\label{finally1}
\eeq
also for $n=3$ we obtain
\beq
{d^2a_3 \over dx^2} =-2\mu a_{5}={a_3 \over5 (x^2-2h)}. 
\label{finally3}
\eeq 
%that can be solved with a solution of the form $a_3=\sum_{n=0}^{\infty} \alpha_nx^n $ where
%\beq
%\alpha_2=-{\alpha_0 \over 20h} \q , \q \alpha_3=-{\alpha_1 \over 60h}
%\eeq
%and
%\beq
%\alpha_{n+2}= {5n(n-1)-1 \over 10h(n+2)(n+1)}\alpha_n.
%\e
Differentiating  the second  of equations (\ref{a_2}) twice  and combining the result
 with equation (\ref{finally1}) and (\ref{a_2}) we find  the
following differential equation
\beq
3(x^2-2h){d^2a_3 \over dx^2}+12x{da_3 \over dx}+2(2+\mu)a_3=0
\label{finally4}
\eeq
Combining (\ref{finally3}) and (\ref{finally4}) we can eliminate the second order derivatives 
and we obtain the following first order differential equation
\beq
{da_3 \over dx}=-{2(2+\mu)+3/5 \over 12x}a_3=-{\gamma \over x}a_3
\label{last}
\eeq
where $\gamma={2(2+\mu)+3/5 \over 12}$ and the general solution is   $a_3=Cx^{-\gamma}$ and 
satisfies equation (\ref{finally3}) if and only if $C=0$. Consequently $a_3=0$ and hence 
 $a_{2n+1}=0$ for ever integer $n$. Therefore $\xi^1 \equiv 0$.    

Similarly we can apply the same reasoning to prove that $\xi^2\equiv 0$. Indeed it is enough to 
consider again  equations (\ref{recurrence3}-\ref{last}) exchanging $x$ and $y$, replacing $\mu$ with $1/\mu$
and $a_n$ with $b_n$.
Therefore we can conclude that there is no  non-null motion, i.e. $g_{ij}\xi^i\xi^j=0$ consequently,
 by Corollary 1, the geodesic flow is not analytically integrable  and thus the initial anisotropic problems 
does not  have any analytic first integrals other than the Hamiltonian. This completes the proof of Theorem 2.

%%%%%%%%%%%%%%%%%%%%%%%%%%%%%%%%%%%%%%%%%%
\ack
%%%%%%%%%%%%%%%%%%%%%%%%%%%%%%%%%%%%%%%%%%
The author is grateful to Scuola Normale Superiore and Centro di Ricerca Matematica Ennio De Giorgi for hospitality. This
work was partially supported by an ESF-PRODYN grant and a Howard E. Petch Research Scholarship.

%%%%%%%%%%%%%%%%%%%%%%%%%%%%%%%%%%%%%

\end{document}